\documentclass[conference]{IEEEtran}
\usepackage{amsmath}
\usepackage{latexsym}
\usepackage{graphicx}
\usepackage{bbding}
\usepackage{indentfirst}
\usepackage{setspace}
\usepackage{float}
\usepackage{epstopdf}
\usepackage{amssymb}
\usepackage{amsfonts}
\usepackage{enumerate}
\usepackage{multicol}
\usepackage{color}
\usepackage{slashbox}
\usepackage{bm}
\usepackage{amssymb}
\usepackage{stfloats}
\usepackage{algorithm,algpseudocode,float}
\usepackage{lipsum}
\IEEEoverridecommandlockouts
\allowdisplaybreaks[4]

\makeatletter
\newenvironment{breakablealgorithm}
  {% \begin{breakablealgorithm}
   \begin{center}
     \refstepcounter{algorithm}% New algorithm
     \hrule height.8pt depth0pt \kern2pt% \@fs@pre for \@fs@ruled
     \renewcommand{\caption}[2][\relax]{% Make a new \caption
       {\raggedright\textbf{\ALG@name~\thealgorithm} ##2\par}%
       \ifx\relax##1\relax % #1 is \relax
         \addcontentsline{loa}{algorithm}{\protect\numberline{\thealgorithm}##2}%
       \else % #1 is not \relax
         \addcontentsline{loa}{algorithm}{\protect\numberline{\thealgorithm}##1}%
       \fi
       \kern2pt\hrule\kern2pt
     }
  }{% \end{breakablealgorithm}
     \kern2pt\hrule\relax% \@fs@post for \@fs@ruled
   \end{center}
  }
\makeatother

\begin{document}
\title{A Bayesian Tensor Approach to Enable RIS for 6G Massive Unsourced Random Access}

\author{\IEEEauthorblockN{Xiaodan Shao$\IEEEauthorrefmark{1}\IEEEauthorrefmark{2}\IEEEauthorrefmark{3}$, Lei Cheng$\IEEEauthorrefmark{1}$, Xiaoming Chen$\IEEEauthorrefmark{1}\IEEEauthorrefmark{2}\IEEEauthorrefmark{3}$, Chongwen Huang$\IEEEauthorrefmark{1}\IEEEauthorrefmark{2}\IEEEauthorrefmark{3}$, Derrick Wing Kwan Ng$\IEEEauthorrefmark{4}$}
\IEEEauthorblockA{$\IEEEauthorrefmark{1}$College of Information Science and Electronic Engineering, Zhejiang University, Hangzhou, China\\
$\IEEEauthorrefmark{2}$International Joint Innovation Center, Zhejiang University, Haining, China \\
$\IEEEauthorrefmark{3}$Zhejiang Provincial Key Laboratory of Info. Proc., Commun. \& Netw. (IPCAN), Hangzhou, China\\
$\IEEEauthorrefmark{4}$School of Electrical Engineering and Telecommunications, University of New South Wales, Sydney, Australia\\
E-mails: \{shaoxiaodan, lei\_cheng, chen\_xiaoming, chongwenhuang\}@zju.edu.cn, w.k.ng@unsw.edu.au
\thanks{The work of L. Cheng was supported in part by the NSFC under Grant 62001309, in part by the Guangdong Basic and Applied Basic Research Foundation under Grant 2019A1515111140. The work of X. Chen was supported by the Zhejiang Provincial Natural Science Foundation of China under Grant LR20F010002 and the Natural Science Foundation of China under Grants 61871344.
The work of Prof. Huang was supported by the National Natural Science Foundation of China under Grant 62101492,
and Fundamental Research Funds for the Central Universities under  Grant 2021FZZX001-21.}
}}\maketitle

\author{\authorblockN{
Xiaodan Shao, \IEEEmembership{Student Member, IEEE}, Lei Cheng,
Xiaoming Chen, \IEEEmembership{Senior Member, IEEE}, Chongwen Huang, \IEEEmembership{Member, IEEE}, and Derrick Wing Kwan Ng, \IEEEmembership{Fellow, IEEE}
}}\maketitle

\begin{abstract}
This paper investigates the problem of joint massive devices separation and channel estimation for a reconfigurable intelligent surface (RIS)-aided unsourced random access (URA) scheme in the sixth-generation (6G) wireless networks. In particular, by associating the data sequences to a rank-one tensor and exploiting the angular sparsity of the channel, the detection problem is cast as a high-order coupled tensor decomposition problem.
However, the coupling among multiple devices to RIS (device-RIS) channels together with their sparse structure make the problem intractable. By devising novel priors to incorporate problem structures, we design a novel probabilistic model to capture both the element-wise sparsity from the angular channel model and the low rank property due to the sporadic nature of URA. Based on the this probabilistic model, we develop a coupled tensor-based automatic detection (CTAD) algorithm under the framework of variational inference with fast convergence and low computational complexity. Moreover, the proposed algorithm can automatically learn the number of active devices and thus effectively avoid noise overfitting.
Extensive simulation results confirm the effectiveness and improvements of the proposed URA algorithm in large-scale RIS regime.
\end{abstract}

\IEEEpeerreviewmaketitle

\section{Introduction}
A typical massive machine-type communication (mMTC) scenario in 6G wireless networks consists of a large number of low-cost devices with sporadic traffics, which only a small fraction of devices are active concurrently.
Under such a setting, one key challenge lies in how to jointly identify the randomly active devices and to estimate their channels in a fast and accurate way \cite{liangAMP}-\cite{unified}.
To overcome this challenge, unsourced random access (URA) was proposed in \cite{PersUra}, which is independent of the total number of devices and depends only on the cardinality of a random active device set. As a result, URA has drawn much research interests due to its effectiveness \cite{wuyongUra,unsourced}. Despite the advancement of URA, its performance is heavily determined by the strength of received signals at the base station (BS) which depends on the channel quality. In this context, emerging wireless technologies for manipulating the wireless channels are desired to unlock the potential of massive URA.

Recently, reconfigurable intelligent surface (RIS), as a promising technology of 6G wireless networks, has been proposed to enhance the received signal quality at desired receivers \cite{huang1}-\cite{huang3}.
The performance gain in RIS-aided communication systems relies  critically on the availability of CSI. Yet, its acquisition is quite challenging in practice due to the passive nature of RIS. In fact, without equipping active radio frequency chains, RIS can neither transmit nor receive pilot signals, thus it is challenging to estimate the channels of RIS to the BS
and devices to RIS separately. Instead, the concatenated device-RIS-BS channels usually are estimated based on the pilot sequences sent from the devices \cite{RISsparse, shi}.

Although various approaches have been proposed in the literature for channel estimation in RIS-aided systems, e.g., \cite{rui, huang2},
the pilot signaling overhead still scales with the product of the number of RIS reflecting elements and devices, which are prohibitively large in practical scenarios, especially for the case of mMTC. As a remedy, the study of joint device separation and channel estimation with a fewer number of pilots for the RIS-aided URA scheme in 6G wireless networks is desired. In general, the considered problem can be formulated as a novel non-standard coupled tensor decomposition problem with a sparse coupled factor, assuming no knowledge about the tensor rank (i.e., the number of active devices). To this end, there are some existing methods which handle coupled tensor decomposition type problems and determine the number of unknown rank \cite{cheng2}-\cite{cheng1}. Unfortunately, the results from these works cannot be directly applied to the joint device separation and channel estimation problem with complicated coupled structure. To address this issue, this paper proposes a novel algorithm for joint devices separation and channel estimation. The contributions
of this paper are as follows:
\begin{enumerate}
\item This paper proposes a novel two-phase framework for RIS-aided massive URA to jointly estimate the channels and separate the devices with only limited pilot sequences.

\item This paper proposes a more advanced element-wise sparsity and low-rank inducing probabilistic modeling and designs a coupled tensor-based detection algorithm under the Bayesian learning framework with automatic active device number and noise power determination.
\end{enumerate}

\emph{Notations}: We use $\otimes$ to denote the Kronecker product, $\circ$ to denote vector outer product, $\left[\!\left[\cdot\right]\!\right]$ to denote the Kruskal operator. $\mathbb{E}$ denotes the expectation operator. $*$ denotes conjugation. $\cdot$ denotes multiplication. $\diamond$ denotes Khatri-Rao product. $\odot$ denotes Hadamard product. $\mathbf{x}$ is said to follow a normal distribution with mean $\mathbf{u}$ and covariance matrix $\boldsymbol{\Sigma}$ of the form $\mathcal{CN}(\mathbf{x}|\mathbf{u},\boldsymbol{\Sigma})$.
$\mathbf{A}(:,n)$ denotes the $n$th column of the matrix $\mathbf{A}$.
\section{Problem Formulation}
\begin{figure}[t]
\setlength{\abovecaptionskip}{-0.cm}
\setlength{\belowcaptionskip}{0.cm}
  \centering
\includegraphics [width=0.43\textwidth] {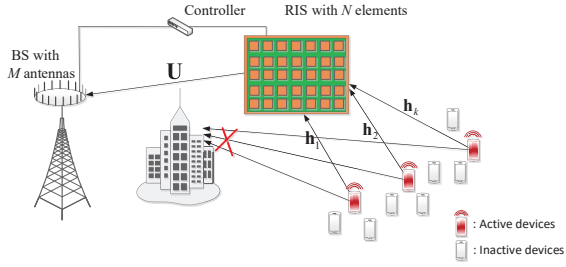}
\caption{RIS-aided massive unsourced random access in 6G wireless networks with blocked direct links.}
\label{sys}
\end{figure}

We consider a RIS-aided 6G wireless network, as illustrated in Fig. \ref{sys}, where a BS equipped with $M$ antennas serves massive single-antenna devices. In order to reduce the access latency and the required signaling overhead in the context of massive devices, the URA scheme is employed in 6G wireless networks. Herein, a RIS consisting of $N$ passive reflecting elements is deployed to enhance the devices' communication performance. The passive reflecting elements of the RIS are arranged as an uniform rectangular array with $N\triangleq N_1\times N_2$. Although the number of potential devices $\bar{K}$ in 6G wireless networks is numerous, only $K_a \ll \bar{K}$ devices are active concurrently at a given slot \cite{liangAMP}. It is assumed that the direct links between active devices and the BS are obstructed \cite{shi} and thus calls for a need to enhance the quality of communications via deploying RIS \cite{kwan3,kwan5,sen}. The channels are assumed to be invariant in a given time slot. In particular, $\mathbf{h}_k\in \mathbb{C}^{N}$ is the channel vector between the RIS and the $k$th device; $\mathbf{U}\in \mathbb{C}^{M\times N}$ is the channel matrix between the BS and the RIS.
%\begin{figure}[t]
%\begin{minipage}[t]{0.43\linewidth}
%\centering
%\includegraphics[width=1.9in]{sys.eps}
%\caption{RIS-aided massive unsourced random access in 6G wireless networks with blocked direct links.}
%\label{sys}
%\end{minipage}%
%\hspace{0.2in}
%\begin{minipage}[t]{0.477\linewidth}
%\centering
%\includegraphics[width=1.9in]{URA.eps}
%\caption{High-level description of the proposed RIS-based URA scheme, where MSG $k$ denotes the message of the $k$th device.}
%\label{URA}
%\end{minipage}%
%\end{figure}
\subsection{Virtual Channel Representation}
Since the RIS is usually mounted at some tall buildings, it is expected that there are only limited scatters around the RIS. This suggests the adoption of the sparse angular channel model \cite{RISsparse}. Specifically, channel $\mathbf{h}_k$ can be modeled as
$
\mathbf{h}_k=\sqrt{\varepsilon_k}\sum_{i=1}^{I_k}\epsilon_i\mathbf{a}_\mathrm{R}(\phi_i,\sigma_i)$,
%\begin{align}\label{angu}
%\mathbf{h}_k=\sqrt{\varepsilon_k}\sum_{i=1}^{I_k}\epsilon_i\mathbf{a}_\mathrm{R}(\phi_i,\sigma_i),
%\end{align}
where $
\mathbf{a}_\mathrm{R}(\phi_i,\sigma_i)=\boldsymbol{\varphi}_{N_2}(-\cos(\sigma_i)\cos(\phi_i))\otimes \boldsymbol{\varphi}_{N_1}(\cos(\sigma_i)\sin(\phi_i))$
with
$
\boldsymbol{\varphi}_{N_p}(x)\triangleq \frac{1}{\sqrt{N_p}}\left[1,\exp^{-j\frac{2\pi}{\varrho}dx},\cdots,\exp^{-j\frac{2\pi}{\varrho}d(N_p-1)x}\right]^T, \forall p=1,2
$;
$N_1$ and $N_2$ denote the length and the width of the rectangular array of RIS; $\varrho$ denotes the wavelength of carrier frequency; $\epsilon_i$ denotes the complex-valued channel gain associated with the $i$-th path; $\phi_i$ and $\sigma_i$ are the azimuth and elevation angle-of-departure (AoD) from the RIS respectively; $\varepsilon_k$ denotes the large-scale path gain for the channel between the RIS and the $k$-th device, and $I_k$ denotes the number of paths between the $k$-th device and the RIS.
Following the grid-based scheme in \cite{RISsparse}, the representation of channel vector $\mathbf{h}_k$ can be further simplified. In particular, two sampling grids $\boldsymbol{\nu}=[\nu_1,\cdots,\nu_{{N_1}^{'}}]^T$, with length ${N_1}^{'}\geq N_1$,
and $\boldsymbol{\varsigma }=[\varsigma_1,\cdots,\varsigma_{{N_2}^{'}}]^T$, with length ${N_2}^{'}\geq N_2$, are employed such that $\mathbf{h}_k$ can be represented as
\begin{align}\label{angu3}
&\mathbf{h}_k=\mathbf{A}_{\mathrm{R}}\boldsymbol{\lambda}_k,
\end{align}
where
$
\mathbf{A}_{\mathrm{R}}\! =\! \left[\boldsymbol{\varphi}_{N_1}(\nu_1),\cdots,\boldsymbol{\varphi}_{N_1}(\nu_{{N_1}^{'}})\right]\!\otimes \! \left[\boldsymbol{\varphi}_{N_2}(\varsigma_1),\cdots,\boldsymbol{\varphi}_{N_2}(\varsigma_{{N_2}^{'}})\right]\in \mathbb{C}^{N \times N_1' N_2'}$,
and $\boldsymbol{\lambda}_k\in \mathbb{C}^{N_1' N_2'}$ represents the channel coefficients of $\mathbf{h}_k$ in the angular domain. Since the number of paths is usually limited, $\boldsymbol{\lambda}_k $ is essentially sparse, with each nonzero value being $\sqrt{\varepsilon_k} \epsilon_i$.

\subsection{RIS-Aided URA Scheme}
In this subsection, we propose a novel two-phase framework of joint device separation and channel estimation for RIS-aided URA scheme. In the first phase, the channel $\mathbf{U}$ from the RIS to the BS is estimated. Specifically, only one selected active device (without loss of generality, device 1) transmits its pilot sequence to the BS through the RIS. The received signal $\mathbf{Y}$ at the BS can be expressed as
\begin{align}\label{rece}
\mathbf{Y}=\mathbf{U}\mathbf{F}+\mathbf{W},
\end{align}
with
$
\mathbf{F}=(\mathbf{V}\odot(\mathbf{h}_1\mathbf{g}_1^T))$,
where $\mathbf{g}_1\in \mathbb{C}^{t_p}$ is the pilot sequence from the device 1 with $t_p$ being the length of pilot sequence; $\mathbf{W}$ is the additive white Gaussian noise; $\mathbf{V}=[\mathbf{v}_1,\cdots, \mathbf{v}_{t_p}]\in \mathbb{C}^{N\times t_p}$ is the reflection beamforming matrix adopted at RIS, in which $\mathbf{v}_t=\left[\alpha_{1,t}\exp(j\theta_{1,t}),\cdots,\alpha_{N,t}\exp(j\theta_{N,t})\right]^T \in \mathbb{C}^{N}, t=1,\cdots,t_p$, is the reflection beamforming vector introduced by the RIS with the ``on/off" state $\alpha_{n,l}\in\{0,1 \}$ and the phase shifts $\theta_{n,l}\in[0,2\pi]$.
It is clear that matrix $\mathbf{F}$ has the same sparsity pattern as the reflect beamforming matrix $\mathbf{V}$ due to the Hadamard product. To impose the sparse structure on $\mathbf{F}$, we design matrix $\mathbf{V}$ as follows: the ``on/off" state $\alpha_{n,t}$ $\sim$ Bernoulli $(a_{n,t}, \jmath)$ independently, where $\jmath$ is the probability of ``on" state;
the phase shifts $\theta_{n,t}$ $\sim$ uniform $(0, 2\pi)$ independently. By exploiting the sparsity of $\mathbf{F}$, the BiG-AMP can be adopted to approximately compute the minimum mean-squared error estimate of $\mathbf{U}$. Details of the BiG-AMP algorithm can be found in \cite{bigamp}.

\begin{figure*}[t]
\setlength{\abovecaptionskip}{-0.cm}
\setlength{\belowcaptionskip}{0.cm}
  \centering
\includegraphics [width=0.67\textwidth]{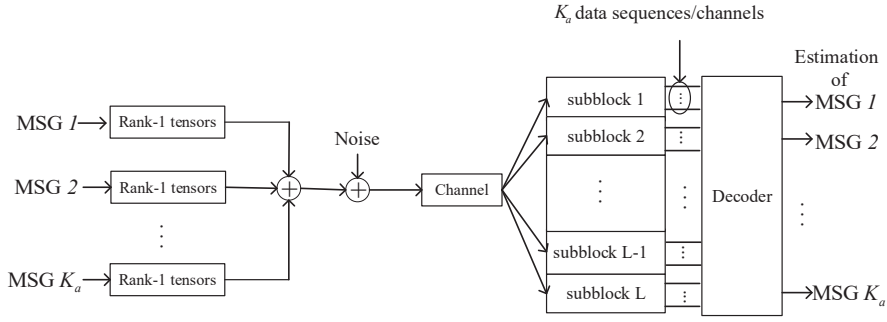}
\caption{High-level description of the proposed RIS-based URA scheme, where MSG $k$ denotes the message of the $k$th device.}
\label{URA}
\end{figure*}

In the second phase, the BS jointly estimates the device-RIS channel and separates active devices, as shown
in Fig. \ref{URA}. Driven by this demand, this paper adopts a block transmission scheme where the $B$-bit message of each active device is divided into $L$ sub-messages \cite{unsourced}. In particular, the length of the $l$-th sub-message is $B_l$ bits. Then, the collection of the sub-messages transmitted within $L$ subblocks can be recovered jointly. Noting that since each subblock $l$ contains $K_a$ recovered sub-messages, only an instance of the transmitted data can be obtained in each subblock. However, the ultimate goal of the BS is to recover the whole set of $B$-bit messages that were transmitted by all the active devices. In this context, a decoder \cite{unsourced} for decoding binary messages among different subblocks $l$ of all the $K_a$ active devices is needed.

Following the URA paradigm \cite{unsourced}, we assume that all the devices in arbitrary subblock $l$ exploit the same constellation  $\wp=\{\mathbf{c}_1,\mathbf{c}_2,\cdots,\mathbf{c}_{2^{B_l}}\}$ containing $2^{B_l}$ elements. Clearly, if devices want to make themselves be identified, they need to embed their identity (ID) information into the transmitted message. Considering the angular domain channel in \eqref{angu3} and the RIS to the BS reflecting channel $\mathbf{U}$, the received signals of the $l$th subblock at the BS can be expressed as
\begin{align}
\label{eqinter5}
\mathbf{Y}_l=\sum_{k=1}^{K_a}{\mathbf{U}}\text{diag}(\mathbf{A}_\mathrm{R}\boldsymbol{\lambda}_k)\mathbf{v}_l\mathbf{s}_{k,l}+\mathbf{W}_l,
\end{align}
where $\mathrm{diag}(\mathbf{x})$ denotes a diagonal matrix with the diagonal entries specified by $\mathbf{x}$, and $\mathbf{s}_{k,l}\in\wp \subset\mathbb{C}^{\tau}$ is the transmitted sequence of complex baseband symbols from the $k$th device over $\tau$
channel uses. $\mathbf{W}_l$ is the additive white Gaussian noise. Herein, the reflection beamforming vector $\mathbf{v}_l$ is constant during the $l$-th subblock and varies from one subblock to another subblock.
Let $\mathbf{y}_l=\mathrm{vec}(\mathbf{Y}_l)$ and $\mathbf{w}_l=\mathrm{vec}(\mathrm{W}_l)$ denote the vectorized versions of $\mathbf{Y}_l$ and $\mathbf{W}_l$, respectively. Then, we rearrange  $\mathbf{U}\text{diag}(\mathbf{A}_{\mathrm{R}}\boldsymbol{\lambda}_k)\mathbf{v}_l$ which gives
$
\mathbf{U}\text{diag}(\mathbf{A}_{\mathrm{R}}\boldsymbol{\lambda}_k)\mathbf{v}_l\!\!=\!\!\sum_{n=1}^{N}\!\mathbf{U}(:,n)\{\mathbf{A}_{\mathrm{R}}\boldsymbol{\lambda}_k\} (n)\mathbf{v}_l(n)\!\!=\!\!\hat{\mathbf{V}}_l\mathbf{A}_{\mathrm{R}}\boldsymbol{\lambda}_k,
$
%\begin{align}
%\label{1}
%\!\!\!\!\!\!\!\!\mathbf{U}\text{diag}(\mathbf{A}_{\mathrm{R}}\boldsymbol{\lambda}_k)\mathbf{v}_l\!\!=\!\!\sum_{n=1}^{N}\!\mathbf{U}(:,n)\{\mathbf{A}_{\mathrm{R}}\boldsymbol{\lambda}_k\} (n)\mathbf{v}_l(n)\!\!=\!\!\hat{\mathbf{V}}_l\mathbf{A}_{\mathrm{R}}\boldsymbol{\lambda}_k,
%\end{align}
where $\hat{\mathbf{V}}_l=[\mathbf{v}_l(1)\mathbf{U}(:,1),\mathbf{v}_l(2)\mathbf{U}(:,2),\cdots,\mathbf{v}_l(N)\mathbf{U}(:,N)]\in \mathbb{C}^{M\times N}$. Note that the phase shifts of $\mathbf{v}_l$ can be set to any value in the range of $[0, 2\pi]$. Then, \eqref{eqinter5} can be equivalently rewritten as
\begin{align}
\label{s5}
\mathbf{y}_l=\sum_{k=1}^{K_a}\mathbf{s}_{k,l}\otimes\mathbf{P}_l\boldsymbol{\lambda}_k+\mathbf{w}_l,
\end{align}
with $\mathbf{P}_l=\hat{\mathbf{V}}_l\mathbf{A}_\mathrm{R}\in \mathbb{C}^{M\times N}$.
Herein, the constellation $\wp$ can be structured according to the tensor decomposition format \cite{newT}. Specifically, it is assumed that the channel use, i.e., $\tau$, can be factorized as $\tau=\Pi_{i}^{d}\tau_i$ for some $d\geq2$ and $\tau_i\geq 2, \forall i $. Subsequently, the complex symbols transmitted by the $k$th device at the $l$th subblock, denoted by $\mathbf{s}_{k,l} \in \mathbb{C}^{\tau}$, can be rewritten as the vector representation of a rank-$1$ tensor $\mathcal{S}_{k,l} \in \mathbb{C}^{\tau_1\times\cdots\times \tau_d}$ of dimensions $\tau_1,\cdots,\tau_d$, that is
%\begin{align}\label{cons}
%\mathbf{s}_{k,l}=\mathrm{vec}(\mathcal{S}_{k,l})\in \mathbb{C}^{\Pi_{i}^{d}\tau_i}=\mathbb{C}^{\tau},
%\end{align}
$
\mathbf{s}_{k,l}=\mathrm{vec}(\mathcal{S}_{k,l})\in \mathbb{C}^{\Pi_{i}^{d}\tau_i}=\mathbb{C}^{\tau}$.
Herein, $
\mathcal{S}_{k,l}=\mathbf{x}_{1,k,l}\circ \mathbf{x}_{2,k,l}\circ \cdots \circ \mathbf{x}_{d,k,l}, \forall k,l$,
where each $\mathbf{x}_{i,k,l}$ is generated from a sub-constellation $\wp_i\subset \mathbb{C}^{\tau_i}$, which is defined as a discrete subset of $\mathbb{C}^{\tau_i}$. In this context, bit information can be mapped to symbol as follows. $B_l$ coded bits are split into $d$ sets of $\{B_{1,1}\},...,\{B_{l,d}\}$ bits, respectively, corresponding to the $d$ tensor dimensions. For the $i$th set, $B_{l,i}$-bits data is mapped to an element of the sub-constellation $\wp_i$. Consequently, we can formulate the design of joint device separation and channel estimation as the following non-convex
optimization problem:
\begin{align}\label{eE2}
&\mathop {\arg\min}\limits_{\mathbf{x}_{i,k,l}\in \wp_i^{\tau_i},\boldsymbol{\lambda}_k\in\mathbb{C}^N}\sum_{l=1}^{L}\left\| \mathcal{Y}_l-\sum_{k=1}^{K_a}\mathbf{x}_{1,k,l}\circ \cdots \circ \mathbf{x}_{d,k,l}\circ (\mathbf{P}_l\boldsymbol{\lambda}_k) \right \|_F^2\nonumber\\
&\mathrm{s. t.}~\|\boldsymbol{\lambda}_k\|_{0}\leq \zeta_s,~~k=1,2,\cdots,K_a,
\end{align}
where $\|\cdot\|_{0}$ denotes the number of nonzero elements of an input vector, $\zeta_s$ is a predefined parameter for imposing the channel sparsity, $\mathcal{Y}_l=\sum_{k=1}^{K_a}\mathbf{x}_{1,k,l}\circ \mathbf{x}_{2,k,l}\circ \mathbf{x}_{3,k,l}\circ \cdots \circ \mathbf{x}_{d,k,l}\circ \mathbf{P}_l\boldsymbol{\lambda}_k+\mathcal{W}_l$ with $\mathcal{W}_l \in \mathbb{C}^{\tau_1\times\cdots\tau_d\times M}$ being the additive white Gaussian noise represented in the tensor space. Note that the tensor rank of noise-free $\mathcal{Y}_l$ is at most $K_a$ \cite{tenrank}.

For the proposed two-phase framework of the RIS-aided URA scheme, the key is to solve problem \eqref{eE2}. However, problem \eqref{eE2} is a non-standard coupled tensor decomposition problem as the sparse profile vector is common and nonlinearly coupled with other parameters in modeling multiple tensors $\mathcal{Y}_l, \forall l$.
Moreover, the number of active devices is random and unknown in practice introducing NP-hardness to the estimation based on tensor data.
These challenges urge the development of a novel efficient algorithm for addressing \eqref{eE2}.

\section{Coupled Tensor-Based Automatic Detection Algorithm}
Note that solving the discrete optimization problem in \eqref{eE2} optimally via an exhaustive search, however, requires $2^{K_aB}$ evaluations of the objective function. To circumvent the complexity issue, we first relax the discrete domain $\mathbf{x}_{i,k,l}\in \wp_i^{\tau_i}$ in \eqref{eE2} to a continuous one $\mathbf{x}_{i,k,l}\in \mathbb{C}^{\tau_i}$. Besides, $K_a$ is unknown for the modeling of all $\mathcal{Y}_l$, $\forall l$, and its estimation has been shown to be NP-hard. To tackle this challenge, an effective way is to introduce two regularization terms that penalizes the model
complexity and avoids possible overfitting of noise:
\begin{align}\label{eT}
&\!\!\mathop {\arg\min}\limits_{{\{\mathbf{X}_l^i\in\mathbb{C}^{\tau_i \times K}\}_{i=1}^{d}}_{l=1}^{L}\mathbf{G}\in\mathbb{C}^{M\times K}}\sum_{l=1}^{L}\left\| \mathcal{Y}_l-\left[\!\!\left[\mathbf{X}_l^1,\cdots,\mathbf{X}_l^d, ~\mathbf{P}_l\mathbf{G}\right]\!\!\right]\right \|_F^2\nonumber\\
&\!\!+\mathop \sum \limits_{k=1}^{K} \gamma_k\left(\sum_{l=1}^{L}\mathop \sum \limits_{i=1}^{d}\mathbf{X}_l^{i}(:,k)^H\mathbf{X}_l^i(:,k)\right)\!+\!\mathop \sum \limits_{k=1}^{K} \eta_k\mathbf{G}(:,k)^H\mathbf{G}(:,k)\nonumber\\
&\!\!\mathrm{s. t.}~\|\mathbf{G}(:,k)\|_{0}\leq \zeta_s,~~k=1,2,\cdots,K,
\end{align}
where $\mathbf{X}_l^i \in\mathbb{C}^{\tau_i \times K}$ with the $k$th column being $\mathbf{x}_{d,k,l}$, $\mathbf{G} \in\mathbb{C}^{M \times K}$ is defined similarly but with the $k$th column being $\boldsymbol{\lambda}_k$.
Herein, we set the column number of all factor matrices $\{\mathbf{X}_l^1, \cdots, \mathbf{X}_l^d, \mathbf{G}\}$ as $K$, which is the maximum possible number of active devices.
In general, if parameters $\gamma_k>0$ and $\eta_k>0$ are sufficiently large after inference, the elements of the $k$th columns in the optimal ${\{\mathbf{X}_l^{i}\}_{l=1}^L}_{i=1}^d$ and $\mathbf{G}$ approach zeros. Then the corresponding column can be pruned out, and the number of remaining columns in each factor matrix gives an estimate of number of active devices. However, the choice of regularization parameters is computationally demanding. Therefore, we develop an intelligent algorithm that can automatically learn both the factor matrices and regularization parameters under the Bayesian learning framework.

\subsection{Element-Wise Sparsity and Low-Rank Inducing Probabilistic Modeling}
Firstly, to apply the Bayesian learning framework, the
probabilistic model, which encodes the knowledge of problem
\eqref{eT}, needs to be established. Therefore, we propose a novel probabilistic model by interpreting each term in problem \eqref{eT} via some probability density functions (pdfs) \cite{cheng3}. We start with the last regularization term in the objective function of \eqref{eT}, which can be modeled as a circularly-symmetric complex Gaussian prior distribution of the columns in matrix $\mathbf{G}$, i.e., $\prod_{k=1}^{K}\mathcal{CN}
\left(\mathbf{G}(:,k)|\mathbf{0},(\eta_k)^{-1}\mathbf{I}\right)$. On the other hand, note that the $l_0$ norm constraint in \eqref{eT} is imposed such that the elements in each column of channel are also sparse. Therefore, by taking both the column-wise sparsity (i.e., low-rank) and the element-wise sparsity structure into account, we propose the following novel prior for $\mathbf{G}$:
\begin{align}\label{pen3}
&p\left(\mathbf{G}|\{\eta_k\}_{k=1}^{K},{\{\boldsymbol{\xi}(n,k)\}_{n=1}^{N}}_{k=1}^{K}\right)\!=\!\prod_{k=1}^{K}\!\left[\mathcal{CN}
\left(\mathbf{G}(:,k)|\mathbf{0},\eta_k^{-1}\mathbf{I}\right)\right.\nonumber\\
&\left.\cdot\prod_{n=1}^{N}\mathcal{CN}\left(\!\mathbf{G}(n,k)|0,\boldsymbol{\xi}(n,k)^{-1}\right)\right],
\end{align}
with $p(\{\eta_k\}_{k=1}^{K}|\boldsymbol{\imath}_{\eta})\!\!=\!\!\prod_{k=1}^{K}\!\!\mathrm{gamma}(\eta_k|\delta,\delta)$, and
$p({\{\boldsymbol{\xi}(n,k)\}_{n=1}^{N}}_{k=1}^{K}|\boldsymbol{\imath}_{\boldsymbol{\xi}})=\prod_{n=1}^{N}\prod_{k=1}^{K}\mathrm{gamma}(\boldsymbol{\xi}(n,k)|\delta,\delta)$,
where $\delta>0$ is a small number that indicates the non-informativeness of the prior model, the natural parameter $\boldsymbol{\imath}_{\eta}=[-\delta\mathbf{1}_{K};(\delta-1)\mathbf{1}_{K}]$, and $\boldsymbol{\imath}_{\boldsymbol{\xi}}=[-\delta\mathbf{1}_{NK};(\delta-1)\mathbf{1}_{NK}]$.
Noting that, after integrating the gamma hyper-prior, the marginal distribution of model parameters in Gaussian-gamma model is a student's $t$ distribution, which is strongly peaked at zero and with heavy tails, thus promoting sparsity.

The proposed prior in \eqref{pen3} is employed to model the sparsity pattern of $\mathbf{G}$. It has a clear physical interpretation. Particularly, $\boldsymbol{\xi}(n,k)^{-1}$ can be interpreted as the power of each element $\mathbf{G}(n,k)$,
while $\eta_k^{-1}$ has a physical interpretation as the power of each column in $\mathbf{G}$.
Therefore, if $\eta_k^{-1}$ is learnt to approach zero, regardless of $\boldsymbol{\xi}(n,k)$, the corresponding columns in the factor matrices play no role in channel modeling and thus can be pruned out by thresholding. Similarly, for a nonzero column ($\eta_{k}^{-1}\neq 0$), $\mathbf{G}(n,k)$ would be shrunk to zero as $\boldsymbol{\xi}(n,k)^{-1}$ goes to zero. Therefore, the proposed prior in \eqref{pen3} can simultaneously promote element-wise sparsity of the channel and low-rank property of tensor data, which mimic the sparsity constraint and the last regularization term in \eqref{eT}.

Similarly, for the regularization term about $\mathbf{X}_l^i$ in problem \eqref{eT}, it can be modeled as a zero-mean circularly-symmetric complex Gaussian prior distribution over the columns of the factor matrices as follows. $
p\left({\{\mathbf{X}_l^i\}_{l=1}^{L}}_{i=1}^{d}|\{\gamma_k\}_{k=1}^{K}\right)=\prod_{l=1}^{L}\prod_{i=1}^{d}\prod_{k=1}^{K}\mathcal{CN}
\left(\mathbf{X}_{l}^i(:,k)|\mathbf{0},\gamma_k^{-1}\mathbf{I}\right)$.
%\begin{align}\label{peX}
%p\left({\{\mathbf{X}_l^i\}_{l=1}^{L}}_{i=1}^{d}|\{\gamma_k\}_{k=1}^{K}\right)=\prod_{l=1}^{L}\prod_{i=1}^{d}\prod_{k=1}^{K}\mathcal{CN}
%\left(\mathbf{X}_{l}^i(:,k)|\mathbf{0},\gamma_k^{-1}\mathbf{I}\right).
%\end{align}
Herein, a gamma distribution is adopted for the penalty parameter $\gamma_k$ to enforce the column-wise sparsity \cite{chengneg}:
$p(\{\gamma_k\}_{k=1}^{K}|\boldsymbol{\imath}_{\gamma})\!=\!\prod_{k=1}^{K}\!\!\mathrm{gamma}(\gamma_k|\delta,\delta)$,
where $\gamma_k^{-1}$ can be interpreted as the power of each column in $\mathbf{X}_l^i$ and $\boldsymbol{\imath}_{\gamma}=[-\delta\mathbf{1}_{K};(\delta-1)\mathbf{1}_{K}]$.

Finally, since the elements of the additive noise $\mathcal{W}_l$ obeys white Gaussian distribution, the sum of the squared error term in problem \eqref{eT} can be interpreted as the negative log of a likelihood function given by
%{\setlength\abovedisplayskip{1pt}
%\setlength\belowdisplayskip{1pt}
%\begin{align}\label{y}
%&p(\{\mathcal{Y}_l\}_{l=1}^{L}|{\{\mathbf{X}_l^i\}_{i=1}^{d}}_{l=1}^{L}, \mathbf{G}, \beta)\propto \exp(-\beta \mathop \sum \limits_{l=1}^{L}\left\| {\mathcal{Y}_l}\right.\nonumber\\
%&\left.-\left[\!\!\left[\mathbf{X}_l^1,~\mathbf{X}_l^2,\cdots,\mathbf{X}_l^d, ~\mathbf{P}_l\mathbf{G}\right]\!\!\right] \right\|_F^2),
%\end{align}}\!\!\!\!\!\!\!\!\!
$p(\{\mathcal{Y}_l\}_{l=1}^{L}|{\{\mathbf{X}_l^i\}_{i=1}^{d}}_{l=1}^{L}, \mathbf{G}, \beta)\propto \exp(-\beta \mathop \sum \limits_{l=1}^{L}\left\| {\mathcal{Y}_l}\right.
\left.-\left[\!\!\left[\mathbf{X}_l^1,~\mathbf{X}_l^2,\cdots,\mathbf{X}_l^d, ~\mathbf{P}_l\mathbf{G}\right]\!\!\right] \right\|_F^2)$,
where the noise precision $\beta$ obeys gamma distribution, i.e., $p(\beta|\boldsymbol{\imath}_{\beta})\propto\beta^{\delta-1}\exp(-\delta\beta)$ with natural parameters $\boldsymbol{\imath}_{\beta}=[-\delta;(\delta-1)]$.
Let
$\boldsymbol{\Theta}\!=\!\!\{{{\{\mathbf{X}_l^i\}_{i=1}^{d}}_{l=1}^{L}}\!, \mathbf{G}\!, \beta\!,{\{\gamma_k\}_{k=1}^{K}}\!,{\{\eta_k\}_{k=1}^{K}}\!,{\{\boldsymbol{\xi}(n,k)\}_{n=1}^{N}}_{k=1}^{K}\!\}$ collects all the unknown random variables.
Given the probabilistic model $p(\boldsymbol{\Theta} ,\{\mathcal{Y}_l\}_{l=1}^L)$, the next goal of Bayesian inference is to learn the model parameters $\boldsymbol{\Theta}$ from the tensor data $\{\mathcal{Y}_l\}_{l=1}^L$, where the posterior probability $p(\boldsymbol{\Theta}|\{\mathcal{Y}_l\}_{l=1}^L)$ is needed to be sought. Noting that maximizing the posterior probability is similar to addressing the problem \eqref{eT}. However, the problem \eqref{eT} cannot learn the regularization parameters.
\subsection{Computation Algorithm for The Inference}
Unfortunately, the proposed probabilistic model is still too complicated to enable an analytically tractable solution since multiple integrations are involved. To address this problem, we adopt the variational inference method which establishes a variational distribution $Q(\boldsymbol{\Theta})$ to approximate the true posterior $p(\boldsymbol{\Theta}|\{\mathcal{Y}_l\}_{l=1}^L)$. To achieve this goal, $Q(\boldsymbol{\Theta})$ is the solution which minimizes the Kullback-Leibler (KL) divergence, i.e., $\mathop\mathrm{minimize} \limits_{Q(\boldsymbol{\Theta})}~\mathrm{KL}(Q(\boldsymbol{\Theta})|p(\boldsymbol{\Theta}|\{\mathcal{Y}_l\}_{l=1}^L))
$.
%\begin{align}\label{kl}
%&\mathop\mathrm{minimize} \limits_{Q(\boldsymbol{\Theta})}~\mathrm{KL}(Q(\boldsymbol{\Theta})|p(\boldsymbol{\Theta}|\{\mathcal{Y}_l\}_{l=1}^L))
%\nonumber\\
%&\triangleq \mathop\mathrm{minimize} \limits_{Q(\boldsymbol{\Theta})}~ -\mathbb{E}_{Q(\boldsymbol{\Theta})}\left\{\ln \frac{p(\boldsymbol{\Theta}|\{\mathcal{Y}_l\}_{l=1}^L)}{Q(\boldsymbol{\Theta})}\right\}.
%\end{align}
To address this problem, mean-field approximation \cite{vi} is widely used to offer a tractable solution, which assumes that the variational pdf can be represented in a fully factorized form, i.e., $Q(\boldsymbol{\Theta})= \mathop \prod \limits_{j=1}^{J}Q({\Theta}_j)$, where ${\Theta}_j\in \boldsymbol{\Theta}$ is part of $\boldsymbol{\Theta}$ with $\bigcup_{j=1}^{J}\boldsymbol{\Theta}_j=\boldsymbol{\Theta}$ and $\bigcap_{j=1}^{J}\boldsymbol{\Theta}_j=\varnothing$, and $J$ is the number of subsets. Under this factorization, each optimal variational pdf $Q^\dagger({\Theta}_j)$ that minimizes the KL divergence can be obtained as follows \cite{vi}
\begin{align}\label{kl3}
Q^\dagger({\Theta}_j)=\frac{\exp(\mathbb{E}_{\prod_{i\neq j}Q({\Theta}_i)}\ln p(\boldsymbol{\Theta},\{\mathcal{Y}_l\}_{l=1}^L))}{\int \mathbb{E}_{\prod_{i\neq j}Q({\Theta}_i)}\ln p(\boldsymbol{\Theta},\{\mathcal{Y}_l\}_{l=1}^L)d\Theta_j}, \forall j.
\end{align}
Using \eqref{kl3}, we can derive the closed-form posterior update for each variational pdf, i.e., $Q^\dagger({\Theta}_j)$.

Since the statistics of each variational pdf rely on other variational pdfs. Therefore, all parameters of variational pdfs need to be updated alternatingly. For clarity, the pseudo-code of the resulting algorithm is outlined in \textbf{Algorithm 1}.
Noting that for a given $\mathbf{s}_{k,l} \in \wp$, $\mathbf{x}_{1,k,l}\circ \cdots \circ \mathbf{x}_{d,k,l}$ cannot be distinguished from $u_1\mathbf{x}_{1,k,l}\circ \cdots \circ u_d \mathbf{x}_{d,k,l}$ with $u_1, \cdots,u_d\in\mathbb{C}$ whenever $\prod_{i=1}^d u_i\!=\!1$. To resolve scalar indeterminacy, each sub-constellation $\wp_i$ can be designed based on a Grassmannian codebook \cite{grcode2}. Then the estimation of $\mathbf{x}_{i,k,l}$ can be searched over the elements of structured Grassmannian constellation and be transformed back to to the discrete domain.
\subsection{Computational Complexity}
Now, we analyze the computational complexity of the proposed CTAD algorithm. For each iteration, the computational complexity is dominated by updating each factor matrix and mainly arises from the matrix multiplication, which is in the order of $\mathcal{O}(L (\mathop \sum\limits_{i=1}^{d}\tau_i+M)K^3+Ld\mathop \prod_{i=1}^{d}M\tau_iK^2+LN^2M)$. Noting that the
computational complexity of single-device demapping is $\mathcal{O}(\tau_i )$, which is independent of data size $B$ and thus is not numerically expensive. It can be seen that the complexity of the proposed CTAD algorithm scales linearly with the subblocks but polynomially with the potential number of active devices. In contrast, the computational complexity of
\begin{breakablealgorithm}
\caption{Coupled Tensor-Based Automatic Detection (CTAD) Algorithm}
\label{alg1}
\begin{algorithmic}[1]
\State  \textbf{Input}: $\{\mathcal{Y}_l\}_{l=1}^L$ and total iterations $T$
\State \textbf{Initialization}: $\mathbf{M}_G^0$, ${\{\mathbf{M}_l^{i,0}\}_{l=1}^{L}}_{i=1}^{d}$, $\boldsymbol{\Sigma}_{G}^0$, ${\{\boldsymbol{\Sigma}_l^{i,t}\}_{l=1}^{L}}_{i=1}^{d}$, $a_{\beta}^0$, $\{a_{\eta_k}^0,a_{\gamma_k}^0\}_{k=1}^K$, and ${\{a_{{\boldsymbol{\xi}}_{n,k}}^0\}_{n=1}^{N}}_{k=1}^{K}$
\For{$t=1 : T$}
\State {\bf{Updates the parameters of $Q(\mathbf{G})^{t+1}$}}:
{\setlength\abovedisplayskip{0.5pt}
\setlength\belowdisplayskip{0.5pt}
\begin{align}\label{al2}
&\!\boldsymbol{\Xi}_l^{t+1}=\left( \frac{a_{\beta}}{b_{\beta}^t}\mathop \odot\limits_{i=1}^{d}\left((\mathbf{M}_l^{i,t})^H\mathbf{M}_l^{i,t}+{\tau_i}\boldsymbol{\Sigma}_l^{i,t}\right)^*\right)^{-1},\\
&\!{\boldsymbol{\Omega}}^{t+1}\!=\!\left(\!\sum_{l=1}^{L}\mathbf{P}_l^H\mathbf{P}_l\otimes(\boldsymbol{\Xi}_l^{t+1})^{-1}\right.\left.+\text{diag}(\mathbf{1}_{N}\!\otimes\![ b_{\eta}/a_{\eta_1}^t \!,\cdots\!,\right.\nonumber\\
&\!b_{\eta}/a_{\eta_K}^t])+\text{diag}( b_{\boldsymbol{\xi}}/{a_{\boldsymbol{\xi}_{1,1}}^t},\left.\cdots,b_{\boldsymbol{\xi}}/{a_{\boldsymbol{\xi}_{1,K}}^t},\cdots,\right.\nonumber\\
&\!\left.b_{\boldsymbol{\xi}}/{a_{\boldsymbol{\xi}_{N,1}}^t},\cdots,b_{\boldsymbol{\xi}}/{a_{\boldsymbol{\xi}_{N,K}}^t})\right)^{-1},\\
&\!\mathbf{u}^{t+1}={\boldsymbol{\Omega}}^{t+1}\mathop \sum \limits_{l=1}^{L}\mathbf{P}_l^H\mathbf{P}_l\otimes(\boldsymbol{\Xi}_l^{t+1
})^{-1}\mathrm{vec}\!\left(\!\left(\!\frac{a_{\beta}}{b_{\beta}^t}(\mathbf{P}_l^H\mathbf{P}_l)^{-1}\right.\right.\nonumber\\
&\!\left.\left.\cdot\mathbf{P}_l^{H}\mathcal{Y}_l(d+1)\left(\mathop \diamond \limits_{i=1}^{d}\mathbf{M}_{l}^{i,t}\right)^*\boldsymbol{\Xi}_l^{t+1
}\right)^H\right),\\
&\!\mathbf{M}_G^{t+1} =\mathrm{reshape}(\mathbf{u}^{t+1},N \times K).
\end{align}}
\State {\bf{Updates the parameters of $Q(\mathbf{X}_l^i)^{t+1}$}}:
{\setlength\abovedisplayskip{0.5pt}
\setlength\belowdisplayskip{0.5pt}
\begin{align}\label{x1}
&\boldsymbol{\Sigma}_l^{i,t+1}=\left(\frac{a_{\beta}}{b_{\beta}^t}\mathop \odot\limits_{j=1,j\neq i}^{d}\left((\mathbf{M}_l^{j,t})^H\mathbf{M}_l^{j,t}+{\tau_j}\boldsymbol{\Sigma}_l^{j,t}\right)^*\right.\nonumber\\
&\left.\odot\left(\mathbf{M}_G^{t,H}\mathbf{P}_l^H\mathbf{P}_l\mathbf{M}_G^t+\mathop\sum \limits_{i=1}^{N}\mathop\sum \limits_{j=1}^{N}\{\mathbf{P}_l^H\mathbf{P}_l\}(i,j)\boldsymbol{\Omega}_{i,j}^{b,t}\right)^*\right.\nonumber\\
&\left.+\text{diag}\left(\frac{b_{\gamma}}{a_{\gamma_1}^t },\cdots,\frac{b_{\gamma}}{a_{\gamma_K}^t }\right)\right)^{-1},\\
&\mathbf{M}_l^{i,t}=\frac{a_{\beta}}{b_{\beta}^t}\mathcal{Y}_l(i)\left(\mathop \diamond \limits \limits_{j=1,j\neq i}^{d}\mathbf{M}_{l}^{i,t}\diamond \mathbf{P}_l\mathbf{M}_G^t\right)^*\boldsymbol{\Sigma}_l^{i,t+1}.
\end{align}}
\State {\bf{Updates the parameters of $Q(\boldsymbol{\xi}(n,k))^{t+1}$}}:
\begin{align}\label{alx}
\!\!\!\!\!\!\!\!\!a_{\boldsymbol{\xi}(n,k)}^{t+1}=\mathbf{M}_{G}^{t+1,*}(n,k)\mathbf{M}_{G}^{t+1}(n,k)+\boldsymbol{\Omega}_{n,n}^{b,t+1}(k,k).
\end{align}
\State {\bf{Updates the parameters of $Q(\eta_{k})^{t+1}$}}:
%\begin{align*}
%a_{\eta_k}^{t+1}
%\end{align*}
%$a_{\eta_k}^{t+1}$ is calculated by \eqref{eeta1}.
{\setlength\abovedisplayskip{0.5pt}
\setlength\belowdisplayskip{0.5pt}
\begin{align}\label{alx1}
\!\!\!\!\!\!\!\!\!a_{\eta_k}^{t+1}&\!=\!(\mathbf{M}_{G}^t(:,k)^{H}\mathbf{M}_{G}(:,k)^t
\!+\![\mathop\sum \limits_{n=1}^{N}\boldsymbol{\Omega}_{n,n}^{b,t+1}](k,k))
\!+\!\delta.
\end{align}}
\State {\bf{Updates the parameters of $Q(\gamma_{k})^{t+1}$}}:
{\setlength\abovedisplayskip{0.5pt}
\setlength\belowdisplayskip{0.5pt}
\begin{align}\label{alx3}
\!\!\!\!\!\!\!\!\!\!\!a_{\gamma_k}^{t+1}&\!=\!\sum_{l=1}^{L}\sum_{i=1}^{d}(\mathbf{M}_{l}^{i,t}(:,k))^H\mathbf{M}_{l}^{i,t}(:,k)\!+\!{\tau_i}\boldsymbol{\Sigma}_{l}^{i,t}(k,k) \!+\!\delta.\!\!\!\!
\end{align}}
\State {\bf{Updates the parameters of $Q(\boldsymbol{\beta})^{t+1}$}}:
{\setlength\abovedisplayskip{0.5pt}
\setlength\belowdisplayskip{0.5pt}
\begin{align}\label{alx5}
&\!\!\!\!\!\!a_{\beta}^{t+1}\!=\!\sum_{l=1}^{L}\!\mathrm{Tr}\left(\mathop \odot \limits _{i=1}^{d}\!\left((\mathbf{M}_l^{i,t})^H\!(\mathbf{M}_l^{i,t})^*\!+\!{\tau_i}\boldsymbol{\Sigma}_l^{i,t}\right)^H\!\! \left(\mathbf{M}_G^{t,H}\mathbf{P}_l^H\right.\right.\nonumber\\
&\!\!\!\!\!\!\left.\left.\cdot\mathbf{P}_l\mathbf{M}_G^t\!+\!\mathop\sum \limits_{i=1}^{N}\mathop\sum \limits_{j=1}^{N}\{\mathbf{P}_l^H\mathbf{P}_l\}(i,j)\boldsymbol{\Omega}_{i,j}^{b,t}\right)\!\!-\!\!\mathbf{P}_l\mathbf{M}_G^t
\left(\mathop \diamond \limits _{i=1}^{d}\mathbf{M}_l^{i,t}\right)^T\!\!\right.\nonumber\\
&\!\!\!\!\!\!\left.\cdot\mathcal{Y}_l(d+1)^H\!-\!\mathcal{Y}_l(d+1)\left(\mathop \diamond \limits _{i=1}^{d}\mathbf{M}_l^{i,t}\right)^*\mathbf{M}_G^{t,H}\mathbf{P}_l^H\right)\nonumber\\
&\!\!\!\!\!+\!\left\|\mathcal{Y}_l(d\!+\!1)\right\|_F^2\!+\!\delta.\!\!
\end{align}}
\EndFor
\State \textbf{Output}: $\mathbf{M}_G^{t+1}$ and ${\{\mathbf{M}_l^{i,t+1}\}_{l=1}^L}_{k=1}^K$
\end{algorithmic}
\end{breakablealgorithm}
the activity detection algorithm in \cite{shi} scales polynomially with the total number of devices, i.e., $\mathcal{O}(\bar{K}^4 )$. Thus,  the proposed CTAD algorithm is computationally efficient.
%\begin{figure*}[t]
%\begin{minipage}[t]{0.277\linewidth}
%\centering
%\includegraphics[width=1.9in]{rank.eps}
%\caption{The estimation accuracy of the number of active devices.}
%\label{rank}
%\end{minipage}%
%\hspace{0.3in}
%\begin{minipage}[t]{0.277\linewidth}
%\centering
%\includegraphics[width=1.9in]{nmseactivity.eps}
%\caption{NMSE versus $L$.}
%\label{nmseactivity}
%\end{minipage}%
%\hspace{0.3in}
%\begin{minipage}[t]{0.277\linewidth}
%\centering
%\includegraphics[width=1.9in]{PERBound.eps}
%\caption{PER versus different $K$.}
%\label{PERBound}
%\end{minipage}%
%\end{figure*}

\section{Numerical Results}
In this section, we present extensive simulation results to validate the effectiveness of the proposed CTAD algorithm in 6G wireless networks.
For the URA scheme, we consider the case that every active device embeds its ID in the payload, in addition to $B = 360$ information bits \cite{unsourced}. It is assumed that there are total of $\bar{K}=4096$ devices in the network, therefore, $B_{\mathrm{ID}} = \log_{2}(K) = 12$ bits are required to encode the device ID.
Hence, each device transmits a total of $B_{\mathrm{tot}} = B_{\mathrm{ID}}+B = 372$ bits.
The device separation error is evaluated by the packet error rate (PER) metric, which is averaged over all active devices. We assume that the entries of RIS-BS channel matrix and the pilot sequences are independent and identically distributed (i.i.d.) zero-mean circularly-symmetric complex Gaussian random variables with unit variance. Every $\epsilon_i$ of the BS-device channel is drawn from complex Gaussian random variables with zero mean and unit variance. The path loss model is set as $\varepsilon_k=l_0(\frac{\bar{d}}{\bar{d}_0})^{\bar{u}}$, where $l_0$ denotes the path loss at the reference distance $\bar{d}_0$. Herein, $\bar{d}_0=1$, $l_0=-30$ dB, and the path loss exponent $\bar{u}$ for device-RIS channel and RIS-BS channel are set as $2$ and $2.5$, respectively. The distance $\bar{d}$ from the $k$th device to RIS is randomly generated from $500$ m to $1000$ m and from the RIS to BS is $100$ m. The angular spread of each subpath is set as $15^{\circ}$.
The over-complete bases $\boldsymbol{\nu}$ and $\boldsymbol{\varsigma}$ in \eqref{angu3} are uniform sampling grids covering $[-1,1]$. The channel estimation accuracy is evaluated in terms of normalized
mean square error (NMSE) given by
$\mathrm{NMSE}=\frac{1}{T_r}\mathop \sum \limits_{i=1}^{T_r}\frac{\left \|{\mathbf{G}}^i-\hat{\mathbf{G}}^i\right \|_F^2}{\left \|{\mathbf{G}}^i\right \|_F^2}$, where $T_r$ is the number of Monte Carlo runs and $\hat{\mathbf{G}}^i$ is the device-RIS channel estimated at the $i$-th run \cite{qiang,wcsp}. The tree-based decoder \cite{unsourced} is employed, where the $B$-bit message of each device is divided into $L$ subblocks of size $B_1,\cdots,B_{L}$ satisfying the following conditions: $\sum_{l}B_l=B$, $B_1=R$, and $B_l<R$ for all $i=2, \cdots, L$.
Herein, all subblocks $i=2,\cdots,L$ are augmented to size $R$ by appending the parity bits which are set to $p_{b}(0,168,270)$. Moreover, we set $R\!=\!270$, $\delta\!=\!10^{-6}$, $d\!=\!2$, $\tau_1\!=\!80$, and $\tau_2\!=\!80$.
\begin{figure}[h]
\setlength{\abovecaptionskip}{-0.cm}
\setlength{\belowcaptionskip}{0.cm}
  \centering
\includegraphics [width=0.42\textwidth] {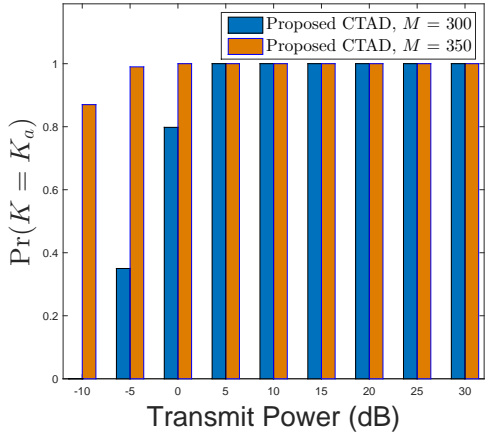}
\caption{The estimation accuracy of active devices number for different transmit power with $K_a\!=\!50$, $N \!=\!500$, and $\bar{K}\!=\!4096$.}
\label{rank}
\end{figure}

When measuring the rank quality of estimation, we consider the mean value of the rank
estimation and probability of the successful recovery in the form of $\mathrm{Pr}(K = K_a)$. Each simulation is repeated 200 times to obtain the averaged result. Fig. \ref{rank} indicates that the success rate of rank estimation is increased as the antenna number $M$ at the BS grows in the lower pilot transmit power region. This is due to the fact that $\mathbf{P}_l$ is equivalent to the measurement matrix in compressed sensing, and the increasing number of BS antennas results in the increment of measurement length for a better observation.
\begin{figure}[h]
\setlength{\abovecaptionskip}{-0.cm}
\setlength{\belowcaptionskip}{0.cm}
  \centering
\includegraphics [width=0.42\textwidth] {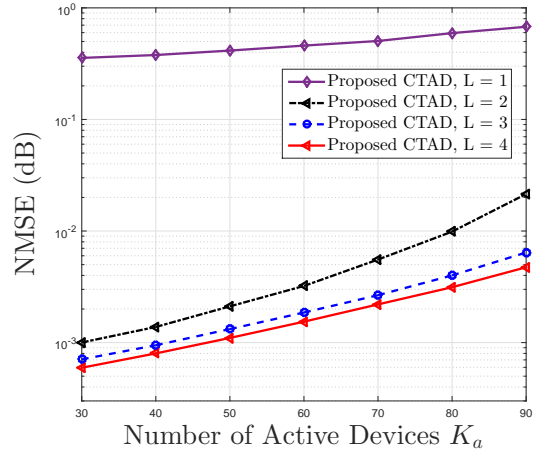}
\caption{NMSE versus $L$ with $M= 256$, $N=350$, transmit power $=15$ dB, and $\bar{K}=4096$.}
\label{nmseactivity}
\end{figure}

In Fig. \ref{nmseactivity}, we focus on the NMSE performance of the proposed CTAD algorithm with different numbers of subblocks $L$ and a fixed rate $B/\tau$. As can be seen from the obtained results, utilizing a single subblock $L=1$ cannot accurately estimate the channel, since $\mathbf{G}$ does not admit a unique solution from the compressed model $\mathbf{P}_l\mathbf{G}$. While the proposed CTAD algorithm is quite accurate for both $L=3, 4$ and the performance degrades with decreasing $L$. From this simulation result, we can see that the coupled tensor factorization criterion in problem \eqref{eT} is critical to the
reconstruction of the channel, since the shared parameter $\mathbf{G}$ in the $L$ fitting terms serves as an anchor to fix the permutation and scaling ambiguities.
\begin{figure}[t]
\setlength{\abovecaptionskip}{-0.cm}
\setlength{\belowcaptionskip}{0.cm}
  \centering
\includegraphics [width=0.42\textwidth] {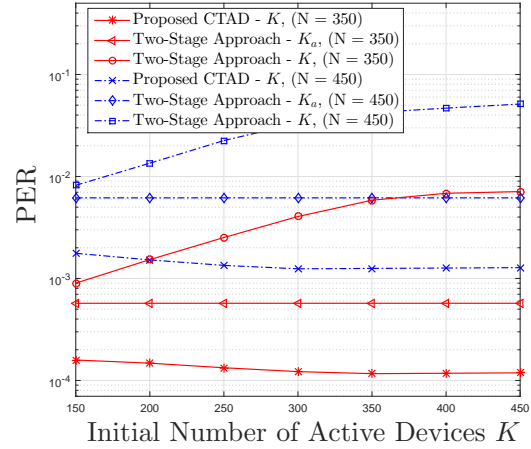}
\caption{PER versus different $K$ with $K_a=50$, $M=256$, $L=3$, transmit power $=10$ dB, and $\bar{K}=4096$.}
\label{PERBound}
\end{figure}

For comparison, we consider a two-stage estimator for addressing the optimization problem in \eqref{eT}: First, the alternating least square (ALS) algorithm \cite{tenrank} is applied to fit the model in \eqref{eT} and $\bar{\mathbf{G}}_l=\mathbf{P}_l\mathbf{G}$ is regarded unknown to estimate. As for the second stage, we compute the minimum $l_1$-norm solutions for the sparse factor matrices $\mathbf{G}$ from the estimation of ${\bar{\mathbf{G}}}_l$ and use the existing convex
optimization solver \cite{convex} to solve it. Since the number of active devices $K_a$ is unknown in practice, we set the initial upper bounds of $K_a$ as $K$. In Fig. \ref{PERBound}, we provide the PER performance curves of the proposed CTAD algorithm under different initial upper bounds of $K_a$, where the genie-aided two-stage approach with exact $K_a$ and the two-stage approach with incorrect active device numbers $K$ are set as benchmarks. First, it can be observed that the proposed CTAD algorithm can offer the best PER results due to its superior capabilities in determining tensor rank, i.e., $K_a$, learning with different $K$. Moreover, the proposed CTAD algorithm is realized through not only element-wise sparsity of channel, but also utilization of the coupled principle for the channels in a clever way. In contrast, the two-stage approach with $K$ overfits the noises heavily and the estimated performance degrades severely in channel estimations.
\section{Conclusion}
In this paper, we established a two-phase framework and then the joint device separation and channel estimation problem was firstly recasted as a coupled high-order tensor problem. Then, a novel probabilistic modeling and an automatic learning algorithm were proposed under the framework of Bayesian inference. The numerical results have shown the remarkable performance of the proposed algorithms.

\end{document}